# Kondo versus Fano in superconducting artificial high-$T_c$ heterostructures

Gaetano Campi[1,2*] Gennady Logvenov[3] Sergio Caprara[4*] Antonio Valletta[5] Antonio Bianconi [1,2*]

[1]*Institute of Crystallography, Italian National Research Council, IC-CNR, Via Salaria Km 29.300, 00015 Roma, Italy email:* gaetano.campi@ic.cnr.it 2
[2]*Rome International Center for Materials Science Superstripes RICMASS, Via dei Sabelli 119A, 00185 Roma, Italy; email* antonio.bianconi@ricmass.eu,
[3]*Max Planck Institute for Solid State Research, Heisenbergstraße 1, 70569 Stuttgart, Germany; email*: g.logvenov@fkf.mpg.de,
[4]*Department of Physics, Sapienza Università di Roma, P.le Aldo Moro, 5, 00185 Roma email: Italy email.* sergio.caprara @uniroma1.it,
[5]*Institute for Microelectronics and Microsystems, Italian National Research Council IMM-CNR, Via del Fosso del Cavaliere, 100, 00133 Roma, Italy. email:* antonio.valletta@cnr.it

*Authors to whom correspondence should be addressed: gaetano.campi@ic.cnr.it; antonio.bianconi@ricmass.eu; sergio.caprara@uniroma1.it; g.logvenov@fkf.mpg.de

**Abstract**

Recently, the quest for high-$T_c$ superconductivity has evolved from the trial-and-error methodology to the growth of nanostructured artificial high-$T_c$ superlattices (AHTS) with tailor-made superconducting functional properties by quantum design. Superlattices are composed of nanoscale superconducting units of modulation doped Mott insulator $La_2CuO_4$ with thickness $L$ intercalated by metallic overdoped $La_{1.55}Sr_{0.45}CuO_4$ and period d. Quantum design based on the multi-gap Bogoliubov theory including spin-orbit coupling (SOC). has been employed for prediction of the amplification of the critical temperature as a function of the conformational parameter L/d. At the top of the superconducting dome, at the *magic* ratio L/d=2/3, the heterostructures are tuned at the Fano-Feshbach resonance and the normal phase exhibits the Planckian T-linear resistivity. Here, we report experimental evidence that the Kondo proximity effect competes with the Fano-Feshbach resonance suppressing $T_c$ on both sides of the superconducting dome. The Kondo proximity effect is expected in electrical resistance of AHTS nanoscale heterostructures following a Kondo universal scaling obtained by numerical renormalization group theory. We show the vanishing Kondo temperature $T_K$ and Kondo scattering amplitude $R_{0K}$ at L/d=2/3, while $T_K$ and $R_{0K}$ increase on the underdoped (L/d>2/3) and overdoped (L/d<2/3) side of the superconducting dome.

**Introduction**

It is known that high-$T_c$ cuprate superconductors can now be engineered by relying on the quantum design of nanoscale artificial high-$T_c$ superlattices (AHTS) of quantum wells [1-3] that capture key features of natural chemically doped cuprates. The growth of these novel nanoscale non-conventional heterostructures has been guided by quantum material design of an artificially modulation-doped [4] correlated electron gas. This design relies on the predictions of the Bianconi-Perali-Valletta (BPV) theory [5-8] describing the amplification of the critical temperature driven by Fano-Feshbach shape resonance [9-12] in two-gap superconductors in the presence of spin-orbit coupling (SOC) at the interfaces in nanostructured materials (NsM) exhibiting quantum size effects.

The three-dimensional AHTS heterostructures are formed by superconductor layers (S) of a stoichiometric modulation-doped Mott insulator $La_2CuO_4$ (LCO), intercalated with potential barriers of normal metal (N) made of chemically overdoped non-superconducting $La_{1.55}Sr_{0.45}CuO_4$ (LSCO) with ten repeats of period variable in the nanoscale range 2.97<d<5.28 nm. The N units plays the role of charge reservoirs and transfer the interface space charge into the superconducting doped Mott insulator units of thickness L. The internal interface electric field at the SNS junctions induces Rashba SOC in the superconducting interface space charge in the S layers which is split into two electronic components by quantum size effects: the lowest subband shows a large cylindrical Fermi surface with *high* Fermi velocity, while the upper subband, exhibits a *low* Fermi velocity and an unconventional extended van Hove singularity generated by SOC at the interface.

Here, we show that at the top of superconducting dome driven by Fano-Feshbach shape resonance the normal phase shows the Planckian T-linear resistivity [13] which competes with Kondo scattering [14-28], resulting in the suppression of $T_c$ on the two sides of the dome. The key result of this work, carried out above the top of the superconducting dome, i.e., in the range 50 K<T<270 K, is the compelling evidence that the temperature dependent sheet resistance of various AHTS in the normal phase evolves as the ratio L/d changes.

It is known that the Kondo proximity effect appears in nanoscale heterostructures of interest in this work. [15]. While the large charge correlation gap due to the local Coulomb repulsion U prohibits tunneling of electrons from the metal into a Mott insulator, the resonant spin flip scattering opens a new channel for tunneling, in this manner the metal 'eats' itself layer by layer into the Mott insulator. This is known as *Kondo proximity effect*, which promotes the formation of metallic interface space charge within the Mott insulator. A peculiar temperature dependence of the Kondo resistivity has been predicted [16] and it was observed in two-component electronic systems, like, e.g., $SrTiO_3/LaTiO_3/SrTiO_3$ heterostructures [17].

Evidence of the Kondo scattering of itinerant electrons by localized electrons has been reported in systems with logarithmic van Hove singularities [18], heavy fermions [19], vanadium dichalcogenides [20], nickelates [21], PrO epitaxial thin films [22], twisted bilayer graphene [23], cuprates [24,25], including Rashba SOC [26]. In our AHTS the superconducting transition temperature is amplified by Fano-Feshbach shape resonance between two superconducting gaps tuned by quantum size effects in the doped Mott insulator nanoscale layers of thickness L due to resonant scattering between closed and open scattering channels. The Fano-Feshbach shape resonance in nanoscale heterostructures has been shown to compete [27] and coexist [28,29] with the Kondo scattering effect.

**Results**

In this work we report experimental evidence of the competition between the Kondo scattering and the Fano-Feshbach shape resonance in AHTS by tuning the chemical potential making quantum superlattices with variable L/d. While the electronic structure of the 2D electron gas at cuprate oxide interfaces with the associated interlayer phase separation has attracted high interest [30,31] we focus on quantum size effects due to confinement of the interphase correlated electron gas in the nanoscale [1]. We have first synthesized AHTS made of quantum wells with period 2.97 nm<d<5.28 nm, tuned at the magic resonant geometry, where pure Mott insulators LCO layers of thickness L, free of chemical substitutions, are intercalated with metallic LSCO layers of thickness W, so that L/d=⅔, where d=L+W. These superlattices exhibit T-linear resistivity in the metallic phase from 50 K to 270 K. By changing the conformational geometry parameter away from the shape resonance, in the range 0.3<L/d<0.9, we observe a Kondo-like temperature dependence in the resistivity, with the Kondo temperature $T_K$ being minimum at L/d=⅔, where the superconducting critical temperature is maximum and the amplitude of the Kondo effect in the resistivity, $R_{0K}$, is minimum.

Therefore, we report compelling evidence that the Kondo scattering competes with the Fano-Feshbach shape resonance.

Nanoscale LSCO/LCO superlattices of alternating overdoped LSCO layers and undoped LCO layers have been grown on a $LaSrAlO_4$ substrate using molecular beam epitaxy (MBE). These superlattices form a 2D electron gas (2DEG) at the interface space charge and exhibit 2D high-$T_c$ superconductivity. The overall superlattice structure has a periodicity represented by the parameter d, as shown in Figure 1a.

Our experimental approach involves the manipulation of the 2DEG interface space charge layer, which extends approximately 2.6 nm into the LCO layer from the LSCO/LCO interface. This layer experiences quantum confinement between two LSCO potential barriers, whose width is denoted as W, within the superlattice. As a result, two artificial subbands emerge, allowing us to modulate their energy splitting or the transparency of the potential barrier by adjusting the thickness ratio L/d. **Figure 1a** depicts a typical LSCO/LCO superlattice with a period of d = 3.96 nm. In this arrangement, the heterostructure consists of five LCO layers (ML), with thickness L = 2.64 nm, while the LSCO layer has a thickness of half ML, W = 1.32 nm. The superconducting critical temperature $T_c$, as a function of L/d assumes a dome shape where, at L/d=⅔, the Fano-Feshbach shape resonance prevails and $T_c$ reaches its maximum values, as shown in Figure 1b.

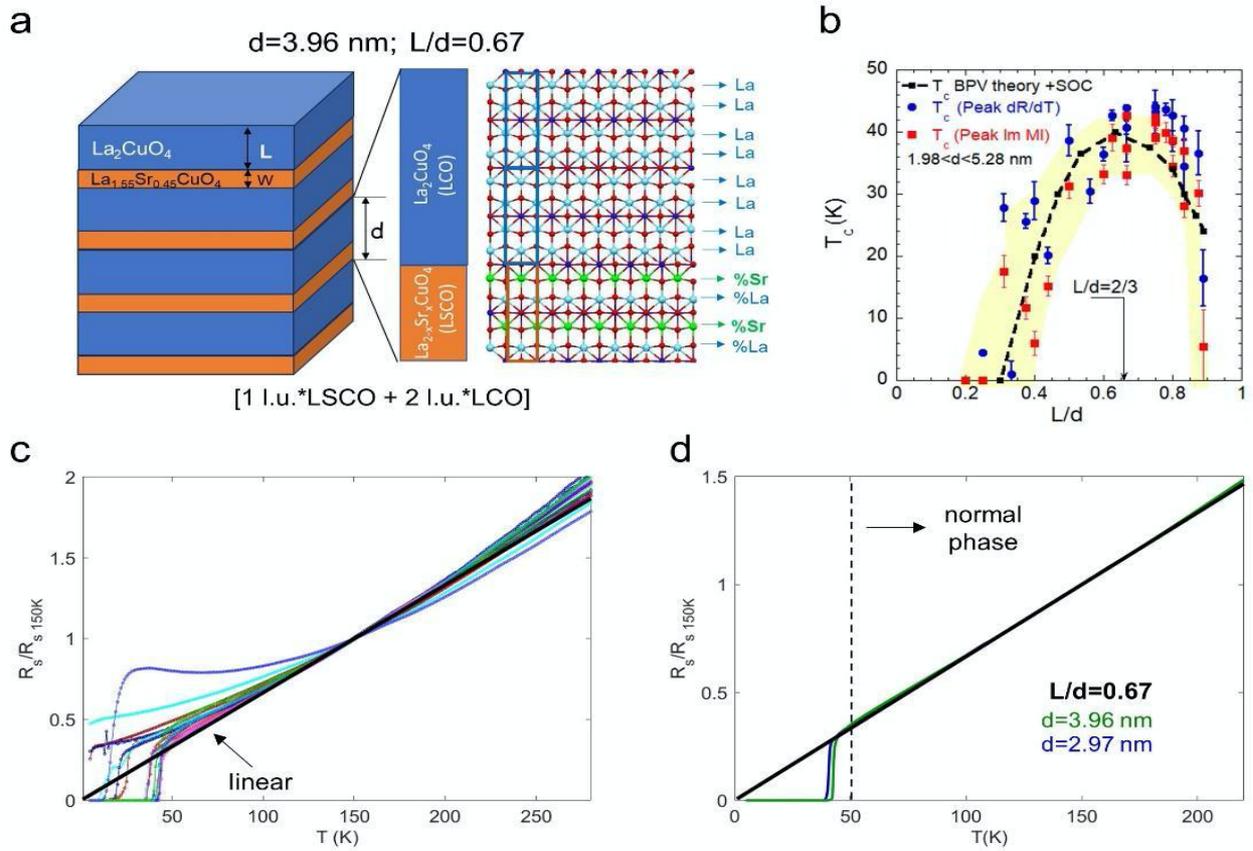

**Figure 1**. **(a)** Pictorial view of a practical realization of the nanoscale AHTS superlattice of quantum wells made of four monolayers ML (L = 2.64 nm) of undoped $La_2CuO_4$ (LCO), electronically doped by the interface space charge, which are intercalated by normal metal units made of two monolayers ML (W = 1.32 nm) of $La_{1.55}Sr_{0.45}CuO_4$ (LSCO) forming a superlattice with a period of d = L + W = 3.96 nm and with conformational parameter L/d = 2/3 giving the optimum critical temperature. **(b)** The superconductivity dome of the critical temperature $T_C$ as a function of L/d in the range of 0.25 < L/d < 0.9. The dashed line represents the predictions by the BPV theory at a Fano-Feshbach shape resonance for the superlattice of quantum wells. We show the $T_c$ determined as maximum of a derivative of the sheet resistance (blue circles) and maximum of the imaginary part of the mutual inductance (red squares). **(c)** Sheet resistance as a function of temperature of several LSCO/LCO superlattices where it is normalized at $R_{s\ 150K}$ which is the resistance measured at 150 K. The samples show the maximum $T_c \approx 43$ K around L/d = 2/3 corresponding to a critical temperature in the optimum doped LSCO. We draw a linear behavior (tick black line) observing how the samples with maximum $T_c$, with L/d=2/3, approach this linear resistivity regime. **(d)** shows the normalized sheet resistance as a function of temperature of two LSCO/LCO superlattices with L/d = 2/3 and different periods d = 3.96 nm and 2.97 nm showing T-linear resistivity in the temperature range 50 K<T<270 K.

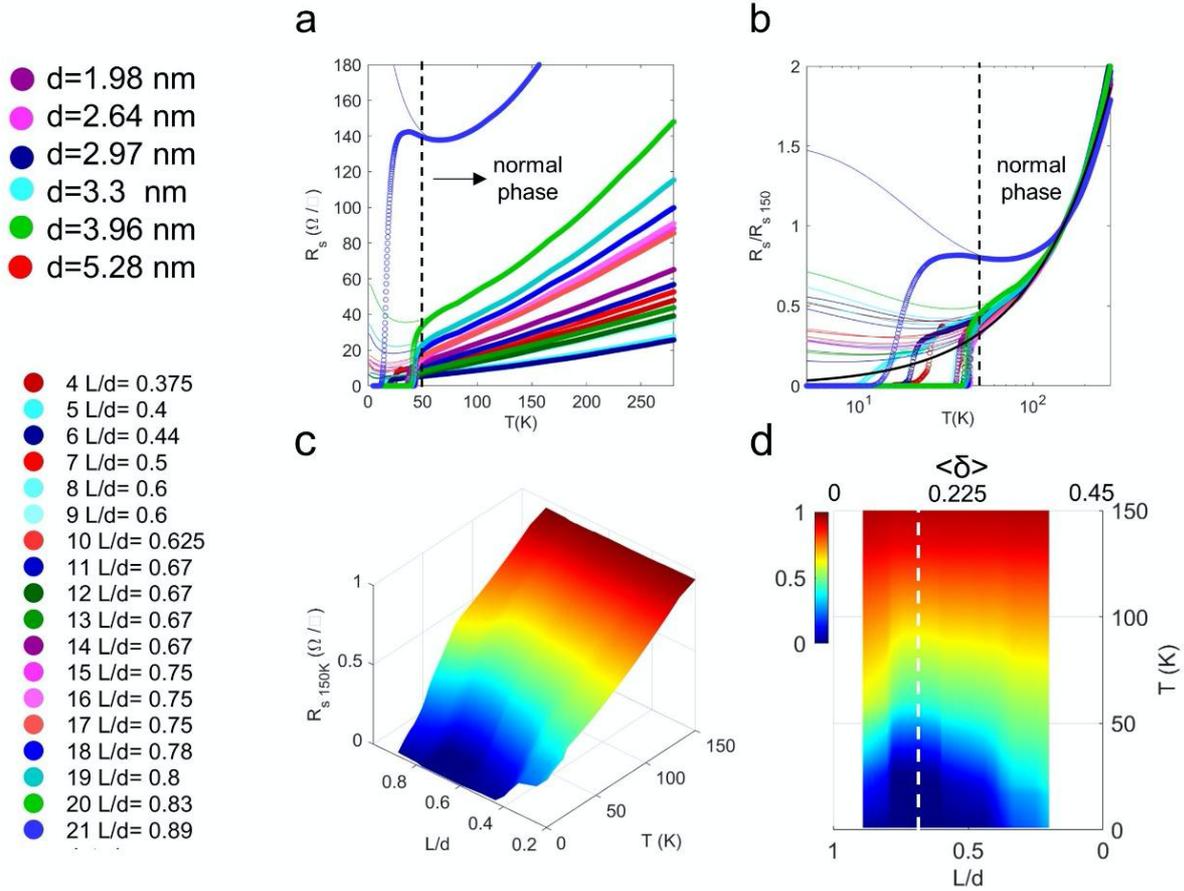

**Figure 2. (a)** (open circles) Sheet resistance as a function of temperature of 18 superlattices samples with different d and L values indicated in the list on the left. The different d-values correspond to the different colors indicated. **(b)** Normalized sheet resistance as a function of temperature alongside the modeled lines through Eq. 1. We show both the (left panel) linear and the (right panel) logarithmic evolution with temperature. **(c)** Color maps of normalized sheet resistance as a function of temperature, the geometrical parameter L/d. In panel **(d)** we show the correspondence between L/d and $<\delta>$ = 0.45 (1-L/d). The white dashed line indicates L/d =2/3 value.

We measured the temperature dependence of the resistance in eighteen LSCO/LCO superlattices with varying values of d, L, and W. The changes in resistivity with temperature, as shown in **Figure 1c**, display curves that both approach and deviate from a linear trend (thick black line). The resistances that most closely follow a linear pattern are reported in **Figure 1d**, where we present the resistance of two significant samples. These samples have the same L/d parameter (L/d = 2/3) but exhibit different d-periodicities. At this L/d value, where the superlattices approach the maximum $T_C$, the FFBR prevails, and a T-linear resistivity is observed in the normal phase.

To investigate the interplay of Fano resonance and Kondo scattering, away from the top of the superconducting dome, we consider here the sheet resistance as a function of temperature and L/d in the measured eighteen LCO/LSCO superlattices listed in Figure 2. These samples have been classified by different d-values, corresponding to different colors and different L/d values. The sheet resistance of all eighteen superlattices, as a function of temperature, are shown in **Figure 2a**, while **Figure 2c** depicts the normalized sheet resistance in units of R(T=150K), in semilogarithmic scale. All R(T) curves have been fitted by using the generalized Kondo equation [16,17]:

$$\frac{R(T)}{R(T=150K)} = r_0 + \frac{T}{150} + AT^2 + BT^5 + \frac{R_{0K}}{\left\{1+\left(2^{\frac{1}{S}}-1\right)\left(\frac{T}{T_K}\right)^2\right\}^S} \qquad (1)$$

for experimental R(T) values measured at T > 50K. The *Kondo temperature* $T_K$ represents a characteristic temperature below which the coupling between the impurity and conduction electrons leads to the screening of the impurity spin. $R_{0K}$ is the amplitude of the Kondo-like contribution, while $r_0$ is the residual contribution at T=0, that represents the baseline resistivity of the material in the absence of Kondo physics including contributions such as impurities and lattice defects. The term $AT^2$ accounts for electron-electron scattering, while $BT^5$ represents phonon scattering. Both $AT^2$ and $BT^5$ terms are non-Kondo contributions. The best fitted curves are represented by continuous lines. The three different representations (Figure 2, panels **a**, and **b**) of the sheet resistance well visualize the fact that the resistivity decreases as the temperature is lowered, reaching a minimum. At even lower temperatures, the electrical resistivity of the system increases (logarithmically, in the original perturbative Kondo treatment, or as a power-law, beyond perturbation theory). We observe how the behavior of the resistance deviates from linearity as the L/d value moves away from 2/3, where the Fano resonance prevails. A phase diagram of normalized sheet resistance as a function of both temperature and the geometrical parameter L/d is given by the 3D color map of Figure 2c. In Figure 2d we visualize a projection of this phase diagram in the T vs. L/d plane, highlighting the correspondence between L/d and the *charge* $\langle\delta\rangle$ = 0.45×(1-L/d).

We have used a least squares fitting algorithm extracting the parameters A, B, $R_{0K}$ and $T_K$. The evolution of these parameters as a function of L/d are shown in Figure 3. In the sheet resistance measured at 150 K in all 18 measured samples with different L/d values, we observe an exponential increase for L/d>2/3, indicating a metal insulator transition triggered by this geometrical conformational parameter. The evolutions of the fitting parameters A, B, $r_0+R_{0K}$ $T_K$ of Eq. 1 are shown in Figure 3b, Figure 3c, Figure 3d and Figure 3e. All parameters show a minimum at L/d =2/3. Finally, in the scatter plot of $T_K$ vs. $R_{0K}$, the positive correlation between these two parameters is made evident in Figure 3f.

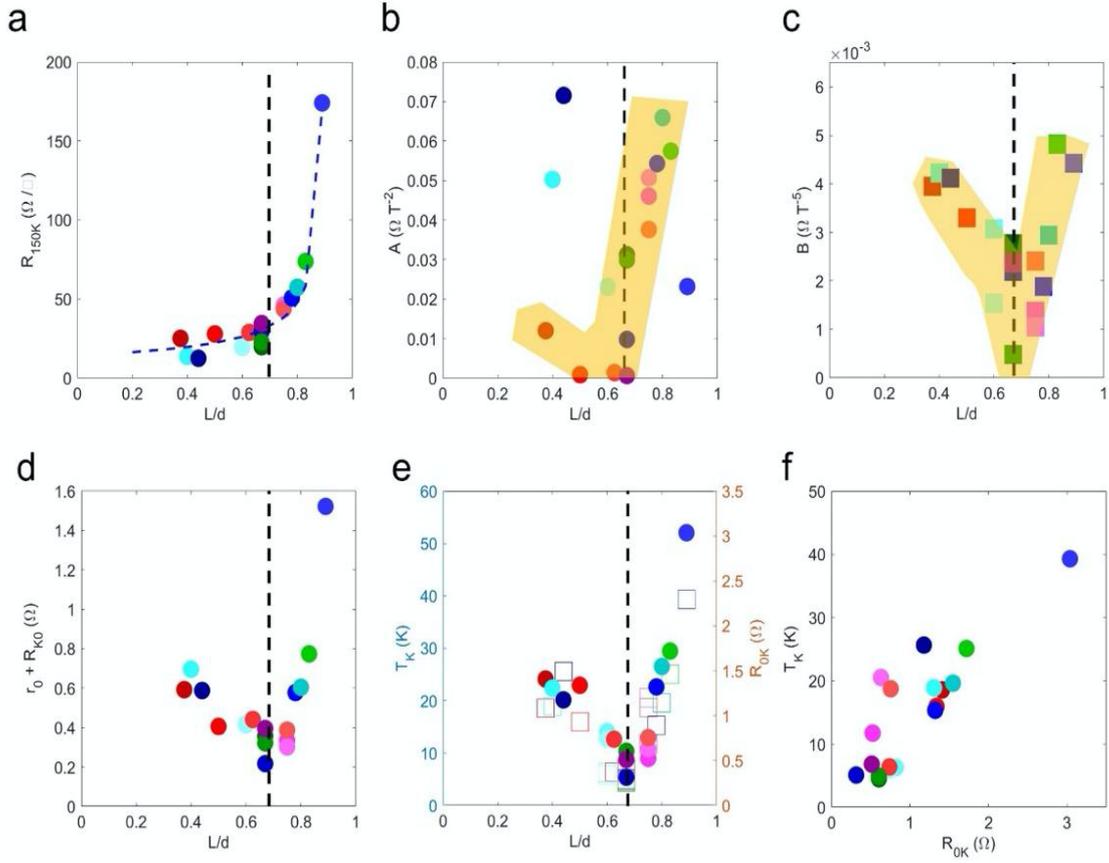

**Figure 3**. **(a)** Sheet resistance measured at 150 K in all 18 measured samples with different L/d values. We observe the exponential increase for L/d>0.667. Evolution of fitting parameters **(b)** A, **(c)** B, **(d)** $r_0+R_{K0}$ **(e)** $T_k$ of Eq. 1. **(f)** Scatter plot of $T_k$ versus $R_{0k}$ showing the positive correlation between these two parameters.

**Materials and Methods+**

*MBE Synthesis of Artificial High-$T_c$ Superlattices:* AHTS based on normal metal LSCO alternating with superconducting space charge layers in LCO thin layers have been synthesized via an ozone-assisted MBE method (DCA Instruments Oy) on LaSrAlO$_4$ (001) substrates (compressive strain for La$_2$CuO$_4$ on LaSrAlO$_4$ is +1.4%). The superlattice growth was controlled by the in-situ reflection high-energy electron diffraction (RHEED). This method is characterized by the sequence of deposition of single atomic layers and minimal kinetic energy of impinging atoms (about 0.1 eV). The substrate temperature, $T_s$, according to radiation pyrometer reading was 650°C, and the chamber pressure p ≈ 1.5 × 10$^{-5}$ Torr of mixed ozone, atomic and molecular oxygen. At the end of the procedure, the samples were cooled down in ozone to $T_s$=220 °C. At this temperature the stoichiometric La$_2$CuO$_4$ layers [44] are free of oxygen interstitials dopants. Then the samples were cooled down in vacuum to avoid doping by oxygen interstitials."

*Resistance measurements:* The temperature dependence of resistance was determined in a four-point van der Pauw configuration with alternative DC current ± 10 μA, in a temperature range from room temperature to 4.2 K (liquid helium). The temperature dependent resistance was measured by using a motorized custom-made dipstick in a transport helium dewar with temperature rate <0.1 K/s.

**Outline**

It has been noticed by many authors that cuprate superconductors exhibiting high-$T_c$ superconductivity show unconventional transport properties in the normal phase above the critical temperature, related with quantum tunneling characterized by *T*-linear resistivity, Planckian limit of the scattering rate assigned to quantum criticality, strong electronic correlation, coexistence of localized and itinerant states [32-43] in a homogeneous strange metal.

In this work we provide further experimental evidence that the Artificial High $T_c$ Superlattices, made of modulation doped quantum wells formed by superconducting (S) layers of chemically stoichiometric Mott insulator $La_2CuO_4$ of thickness on the range of 2-3 nanometers intercalated by a normal metal (M), with the role of charge reservoir and a potential barrier of 500 meV [1] grab the key feature of natural chemical doped $La_{2-x}Sr_xCuO_4$.

We confirm that AHTS exhibit first (i) the *superconducting dome* of the critical temperature versus modulation doping and second (ii) T-linear Planckian sheet resistance around optimum modulation doping reported in reference [1]. These results falsify the decades old paradigm of a homogeneous strange metal near a quantum critical point for the interpretation of the phase diagrams of unconventional high-$T_c$ superconductors. The present results support the physical paradigm describing unconventional high $T_c$ cuprate perovskites characterized by intrinsic functional arrested chemical nanoscale phase separation observed by local and fast x-ray experimental probes [44-50] of the quantum complex matter landscape called "superstripes" [46] or "swiss cheese" [48] scenario where the electron gas is near a BEC-BCS crossover.

The key remarkable result, shedding further light on the mechanism of high $T_c$ superconductivity, is that the proximity Kondo resistance competes with T-linear regime in the normal phase sheet resistance of doped cuprate AHTS. Where the Fano Feshbach resonance between two superconducting gaps determined by quantum size effects of the nanoscale units has been optimized by quantum design at the geometrical conformational parameter of the superlattice, L/d=2/3, at the top of the superconducting dome the Kondo temperature $T_K$ reaches a minimum of the order of 4 Kelvin and the Kondo amplitude $R_{0K}$ vanishes as shown in Fig.3. On both sides of the superconducting dome where the average doping is in the so-called overdoped L/d<2/*3 or* underdoped regime L/d>2/3 the Kondo temperature $T_K$ increases and becomes higher than the superconducting critical temperature. These results are supported by the clear experimental anticorrelation between $T_K$ and $R_K$. Finally it is remarkable that where the resonant quantum tunnelling driven by Fano-Feshbach resonant is dominant also the residual T=0K resistance, the amplitude of the electron-electron Fermi scattering term $T^2$ and the electron-phonon $T^5$ term vanish in the normal phase.

**Authors**

Email : antonio.bianconi@ricmass.eu, g.logvenov@fkf.mpg.de, sergio.caprara@uniroma1.it, antonio.valletta@cnr.it, gaetano.campi@ic.cnr.it

ORCID: Antonio Bianconi 0000-0001-9795-3913; Gennady Logvenov 0000-0003-1986-0249; Sergio Caprara 0000-0001-8041-3232 ; Antonio Valletta 0000-0002-3901-9230; Gaetano Campi 0000-0001-9845-9394

**Author Contributions:**
*Conceptualization,* A.B. and G.L and G.C..; methodology, G.L. and A.B.;
*software,* A.V., G.C. and A.B.; validation, G.L. and A.B.;
*formal analysis,* G.L., G.C. A.B.;
*investigation,* G.L., G.C. A.V. and A.B.;
*sample synthesis,* G.L.;
*data curation,* G.L., G.C. and A.B.;
*writing original draft preparation* A.B., G.C. , G.L. and S.C.;
*writing review and editing*, A.B., G.C.. S.C., ,G.L.;
All authors have read and agreed to the published version of the manuscript

**Funding:** This research received funding from Superstripes onlus.

**Data Availability Statement:** The data that support the findings of this study are available from the corresponding authors (A.B.. G.C.) upon reasonable request.

**Conflicts of Interest:** The authors declare no conflict of interest. The funders had no role in the design of the study; in the collection, analyses, or interpretation of data; in the writing of the manuscript; or in the decision to publish the results